\documentclass[iop,revtex4]{emulateapj}
\usepackage{lineno}

\usepackage{lscape}

\begin{document}

\renewcommand{\topfraction}{1.0}
\renewcommand{\bottomfraction}{1.0}
\renewcommand{\textfraction}{0.0}

\shorttitle{Speckle interferonmetry at SOAR}
\shortauthors{Tokovinin et al.}

\title{Speckle interferometry at SOAR in 2016 and 2017 }

\author{Andrei Tokovinin}
\affil{Cerro Tololo Inter-American Observatory, Casilla 603, La Serena, Chile}
\email{atokovinin@ctio.noao.edu}

\author{Brian D. Mason \& William I. Hartkopf}
\affil{U.S. Naval Observatory, 3450 Massachusetts Ave., Washington, DC, USA}
\email{brian.d.mason@navy.mil}
\author{Rene A. Mendez}
\affil{Universidad de Chile,  Casilla 36-D, Santiago, Chile}
\email{rmendez@u.uchile.cl}
\author{Elliott P. Horch\footnote{Adjunct Astronomer, Lowell Observatory} }
\affil{Department of Physics, Southern Connecticut State University, 501 Crescent Street, New Haven, CT 06515, USA}
\email{horche2@southernct.edu}

\begin{abstract}
The results of speckle interferometric  observations at the 4.1 m SOAR
telescope in  2016 and 2017  are given, totaling 2483  measurements of
1570 resolved  pairs and 609 non-resolutions.  We describe
briefly  recent changes  in the  instrument and  observing  method and
quantify  the   accuracy  of  the  pixel  scale   and  position  angle
calibration.  Comments  are given  on 44 pairs  resolved here  for the
first time. Orbital motion of the newly resolved subsystem BU~83 Aa,Ab
roughly  agrees  with  its  36  year  astrometric  orbit  proposed  by
J.~Dommanget.  Most {\it Tycho} binaries examined here turned out to be spurious.
\end{abstract} 

\section{Introduction}
\label{sec:intro}

We report  here a  large set of  double-star measurements made  at the
4.1 m  Southern  Astrophysical  Research  Telescope  (SOAR)  with  the
speckle camera,  HRCam.  This paper continues the  series published by
\citet[][hereafter   TMH10]{TMH10},  \citet{SAM09},  \citet{Hrt2012a},
\citet{Tok2012a},  \citet{TMH14},  \citet[][hereafter  SOAR14]{TMH15},
and  \citet{SAM15}.  

The  objects  were  selected  mostly  among  nearby  (within  200\,pc)
binaries resolved  by {\it Hipparcos}, continuing  our previous effort
in this direction that  mirrors the effort in recent years at WIYN
\citep{Horch2017}   and  DCT   \citep{Horch2015}  telescopes   on  the
analogous Northern  {\it Hipparcos}  sample.  Similar to  the Northern
program,  the main  goal is  to  identify and  follow {\it  Hipparcos}
binaries that  show relatively fast  orbital motion and that  would be
good candidates  for mass determinations  in the coming years.   It is
especially  important  to obtain  orbital  data  on  these systems  in
advance of  final {\it Gaia}  results, so that  the full power  of the
{\it Gaia} parallaxes  can be brought to bear  on the determination of
the mass sum. We also followed  the fast orbital motion of close pairs
and subsystems  discovered previously at  SOAR or elsewhere,  with the
aim of characterizing  their orbits. These data are  actively used for
orbit  calculation   \citep{Gomez2016,Tok2017c,Mdz2017,Tok2018b,Tok2018c,Mason2018}.   To  provide
additional bright  targets when  observing conditions are  mediocre or
poor, we  observed as  a ``filler'' potentially  interesting neglected
pairs suggested by R.~Gould (private communication) and binaries with known orbits.

\section{Observations}
\label{sec:obs}

\subsection{Instrument}
\label{sec:inst}

The   observations  reported   here  were   obtained  with   the  {\it
  high-resolution camera} (HRCam) -- a fast imager designed to work at
the 4.1 m SOAR  telescope \citep{TC08,Tok2018a}. For practical reasons, the
camera was  mounted on  the SOAR Adaptive  Module \citep[SAM,][]{SAM}.
However, the  laser guide  star of  SAM was not  used (except  in 2016
January)   because it  was not needed  and, moreover,  reduced the
  productivity by adding an overhead.  The deformable mirror of SAM was passively flattened
and  the  images are  seeing-limited.   The  SAM  module contains  the
atmospheric  dispersion corrector  (ADC)  and helps  to calibrate  the
pixel scale  and orientation of HRCam (see  SOAR14).  The transmission
curves of HRCam filters  are given in the instrument manual.\footnote{
  \url{http://www.ctio.noao.edu/soar/sites/default/files/SAM/\-archive/hrcaminst.pdf}}
We  used  mostly  the  Str\"omgren  $y$ filter  (543/22\,nm)  and  the
near-infrared $I$ filter (788/132\,nm).

In 2016 May, at the end of  the run, the Luca-DL detector of the HRCam
failed  after ten  years of  faithful service  and one  repair  by the
vendor  (Andor) during  this period.   In 2016  December, we  used the
Luca-R  camera loaned  by the  STELES instrument  team.  With  a 75-mm
camera lens, the  pixel scale was 14.30 mas.   A similar Luca-R camera
was  also  used  in  2014,  as  described  in  SOAR14.  However,  that
frame-transfer  CCD  had  imperfect  charge  transfer  in  the  column
direction,  leading to  a partial  loss of  resolution. This  time, the
loaned Luca-R camera  was characterized in this respect,  and we found
that it presents  a similar problem, although to  a smaller extent.  A
typical charge  spread along the columns was  found to be from  2 to 3
pixels.  We  did not  account for this  effect in the  data processing
(Section~\ref{sec:dat}), as  in 2014 \citep{TMH14}, but  in some cases
could  reduce its  influence  by  using a  reference  spectrum with  a
comparable  smear. One  consequence  of this  problem  is the  reduced
resolution  in the  vertical (usually  North-South) direction  for the
fainter targets.

Meanwhile, Dr. N.~Law from the University of North Carolina has kindly
loaned us  a better electron  multiplication (EM) CCD camera,  iXon X3
888  (hereafter  iXon-888),   also  manufactured  by  Andor.\footnote{
  \url{http://www.andor.com/cameras/ixon-emccd-camera-series}}   Unlike
Luca, this  is a back-illuminated  EM CCD with a  substantially higher
quantum efficiency  and a deeper  cooling.  The detector  has 1024$^2$
pixels.  Sending  this export-controlled  camera to Chile  took longer
than expected, so it could be used only in 2017.  The larger 13-micron
pixel size  required a change of  the re-imaging lens in  HRCam to one
with  125\,mm focal  length,    resulting in  the  pixel scale  of
  15.75\,mas.  The mechanical structure  was reinforced to hold this
heavier camera. The HRCam PC  computer was also replaced. As the cable
connecting camera to  the computer is short, the  PC was located close
to the  HRCam.  The data acquisition  software was adapted  to the new
detector by R.~Cantarutti.  The minimum exposure time for the standard
200$\times$200  pixel region was  24\,ms, and  this exposure  time was
used  mostly throughout  2017.   With this  exposure  time, the  50-Hz
vibrations, when  present, affect  both resolution and  sensitivity of
HRCam \citep{SAM09,Tok2018a}.   A shorter exposure  of 6.7\,mas is  possible in
the  so-called cropped  sensor  mode.   This mode  was  tested at  the
telescope, but not used because switching between readout modes cannot
be done rapidly.

\begin{figure}
\epsscale{1.1}
\plotone{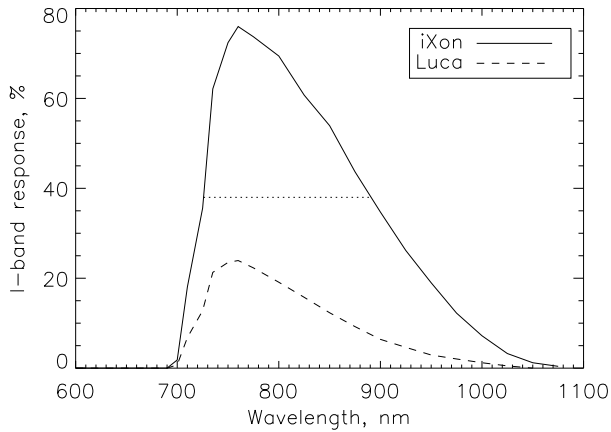}
\caption{Product of the $I$-filter transmission curve and the detector
  quantum efficiency.  The dotted horizontal  line shows the  FWHM for
  the iXon camera.
\label{fig:filt} }
\end{figure}

Figure~\ref{fig:filt}  shows  the   quantum  efficiency  (QE)  of  the
detector multiplied by the transmission of the filter $I$. This filter
cuts  off only  short wavelengths,  so  the bandwidth  depends on  the
detector's response.  The curve for  iXon-888 has a Full Width at Half
Maximum  (FWHM)  of  170\,nm  (from  725\,nm  to  895\,nm),  with  the
effective  wavelength  of  824\,nm.   With such  wide  bandwidth,  the
effective wavelength  depends on the source color;  it is substantially
longer for red  stars.  For comparison, the corresponding  QE curve of
the Luca-DL  camera with the same  filter is also plotted;  it has the
central  wavelength  of  788\,nm  and  a narrower  FWHM  bandwidth  of
132\,nm.

\subsection{Observing procedure}
\label{sec:proc}

In speckle  interferometry, the data  accumulation takes only  a short
time (typically  8\,s). Therefore, the observing  efficiency is mainly
determined by  the telescope  slew and setup  (centering of  the star,
setting  the ADC,  changing  filters).   The standard  procedure  at  SOAR
requires the operator  to select target coordinates from  a list and
to  command the slew  to the  new position.  When the  observing list
contains several hundred targets, this is a labor-intensive task.

To address  this problem, we  developed in 2014 the  speckle observing
tool, written in IDL. In preparation for the run, targets are selected
from  the database  including all  observed objects;  it  contains the
equatorial coordinates, proper  motions, magnitudes, separations, date
of  the  last  observation   at  SOAR,  and  comments.   The  selected
information  is  used  by  the  observing tool  with  a  graphic  user
interface. It displays  an area of the sky  around the selected target
in  the  horizontal (azimuth  and  elevation)  coordinates.  The  next
target  is  selected  by  clicking  on the  display  or  entering  its
number. All previous  observations of the selected object  at SOAR can
be listed on the screen, if desired.  By pressing a button, the target
coordinates for  the current moment (accounting for  the proper motion)
are sent to the SOAR telescope control system (TCS).  The new SOAR TCS
moves  the  telescope  if  the  requested slew  is  less  than  5\degr
(recently extended to 15\degr),  otherwise confirmation of slew by the
telescope operator is  needed.  This tool greatly reduces  the load on
the telescope operators. At the same  time, the target name is sent to
the instrument  software, hence  there is no  need to  type it.
The observer  only has to center  the target and to  select the filter
and the detector parameters.  Use  of this observing tool has improved
the efficiency to the point where 300 targets could be observed in one
winter night.

The choice  of the next target remains  manual, considering priorities
and  variable observing  conditions. For  example, useful  measures of
bright stars can be made through transparent clouds or under very poor
seeing.  So,  the observing program contains  extra ``filler'' targets
for  such  situations.  The  observing  tool  also  helps to  optimize
telescope  slews.    Combination  of  priorities,   diverse  observing
conditions,  slew  and visibility  constraints  defines which  program
stars are actually  observed in each run.  In  2017, most observations
were made remotely from La Serena. This is very convenient, especially
for short observing runs.

\subsection{Observing runs}

The observing  time for this  program was allocated through  NOAO (2.5
nights in 2016,  programs 16A-0005 and 16B-0044, PI  A.T.)  and by the
Chilean National Time Allocation  Committee, CNTAC (4 nights in 2016A,
program CN2016AB-4,  PI R.A.M.).  Some data reported  here (e.g.  on
calibration  binaries) were  also collected  during  observations with
HRCam  for   other  programs  (2.5  nights  in   2015B,  15B-0268,  PI
C.~Brice\~no,  1 night  in 2016A/B,  PI  B.~Pantoja, and  2 nights  in
2016A, PI Ji~Wang, {\it  Kepler-2} follow-up).  Measures and discoveries
resulting from the {\it Kepler-2} (K2) program are included in this paper. In
2017,  two  nights  per   semester  were  assigned  through  NOAO  for
multiple-star observations (PI  A.T., programs 17A-0008 and 17B-0066).
All observations were made by A.T., sharing the allocated time between
 programs  to cover the whole  sky and to  improve temporal cadence
for pairs with fast orbital motion.

\begin{deluxetable}{ l l r r c c } 
\tabletypesize{\scriptsize}    
\tablecaption{Observing runs
\label{tab:runs} }                    
\tablewidth{0pt}     
\tablehead{ \colhead{Run}  &
\colhead{Dates}  &  
\colhead{$\theta_0$} & 
\colhead{Pixel} &
\colhead{$N_{\rm obj}$} &
\colhead{$\beta$} 
\\H
 &   &  \colhead{(deg)} & \colhead{ (mas) } & &
\colhead{(\arcsec)}
}
\startdata
1 & 2016 Jan 16-18   & $-3.00$  & 15.23 & 270 & 0.77 \\ 
2 & 2016 Feb 18-20   & $-2.72$  & 15.23 & 474 & 0.83 \\ 
3 & 2016 May 20-23   & $0.20$   & 15.23 & 315 & 1.15 \\ 
4 & 2016 Dec 12      & $-12.0$  & 14.30 & 41  & 0.64 \\ 
5 & 2016 Dec 15-17   & $0.30$   & 14.30 & 493 & 0.67 \\  
6 & 2017 Apr 13      & $-0.1$   & 15.75 & 152 & 0.82 \\ 
7 & 2017 May 15      & $-0.1$   & 15.75 & 201 & 0.80 \\ 
8 & 2017 Jun 6       & $-0.1$   & 15.75 & 319 & 0.68 \\ 
9 & 2017 Jul 14      & $-0.1$   & 15.75 & 161 & 0.88 \\ 
10& 2017 Aug 7       & $0.1$    & 15.75 & 116 & 1.02 \\ 
11& 2017 Sep 5       & $0.0$    & 15.75 & 275 & 0.73 \\ 
12& 2017 Oct 4       & $-0.15$   & 15.75 & 41  & 0.90  \\   
13 & 2017 Oct 28     & 0.2      & 15.75 & 122 & 0.55 
\enddata
\end{deluxetable}

Table~\ref{tab:runs}  lists   the  observing  runs,   the  calibration
parameters (position angle offset  $\theta_0$ and pixel scale in mas),
and the number  of objects {\it observed for  all programs} covered in
each run. Its last column gives the median FWHM of the re-centered
images, $\beta$, determined during data processing.

{\it Run 1} (2.5 nights in 2016 January) was dedicated to observations
of  young  stars (PI  C.~Brice\~no) in  Orion  and  Chamaeleon,  see
\citet{Cha}.  The image quality (hence sensitivity)  was improved using
the UV  laser, allowing us  to observe stars  of $I \sim 13$  mag with
exposure times of  0.1 or 0.2\,s (not quite  at the diffraction limit,
though, with a median FWHM  of 0\farcs33).  When the main targets were
not visible, double stars were  observed without laser in the standard
speckle mode. The seeing was good during most of this run.

{\it  Run  2}  (2.5  nights   in  2016  February)  was  split  between
observations of multiple stars  (0.5 nights), {\it Hipparcos} binaries
(2  nights), and  the program  of B.~Pantoja  (0.5 night).   It enjoyed
clear skies and  slow wind speed, with  average  seeing.  On the last
night of  the run, the phenomenon  of optical ghosts  was observed, as
described below in Section~\ref{sec:ghosts}.

{\it Run  3} (4  nights in  2016 May) suffered  from poor  weather and
technical problems.  The first night  started with a strong wind and a
high humidity  of 83\%.  The  seeing was extremely poor  (2\arcsec ~to
3\arcsec).  The  telescope was  closed for high  humidity most  of the
night, so only 44 bright stars could be observed.  The following night
was clear, with a strong  wind, poor seeing, and occasional passage of
transparent clouds.   The third night  was lost to clouds.   When some
bright stars  were observed between  the clouds, frequent  failures of
the Luca-DL camera prevented operation.   On the last night of the run
(also  mostly cloudy)  we  replaced  the camera,  but  these data  are
discarded here, being insufficient in  both quality and quantity for a
meaningful analysis.  The  first two nights of this  run were assigned
to  the {\it  Kepler-2}  (K2)  follow-up, which  is  the Yale  program
managed  by  Ji Wang  (Caltech).   He  provided  the list  of  targets
including all known binaries  from the Washington Double Star Catalog,
WDS  \citep{WDS} in  the  K2 fields  (RA  range from  16\,h to  20\,h,
declination from $-31\degr$ ~to $-25$\degr). Pairs wider than 4\arcsec
~could  not be  measured with  HRCam;  however, we  discovered 10  new
components  in known  binaries (see  Section~\ref{sec:new}).  The  two last
nights  (May 22  and 23)  were assigned  to the  program  of R.~Mendez
(CNTAC) on  {\it Hipparcos} binaries.  In  this run, the  new SOAR TCS
was operational, allowing small slews  to be commanded directly by the
speckle observing tool.

{\it Run 5}  (2.5 nights in 2016 December) was preceded  by 2 hours of
engineering observations  on December 12,  treated here as  a separate
run 4.  As  described in Section~\ref{sec:inst}, we used  the loaned Luca-R
camera in these runs.  The sky  was clear, the seeing was average. 

{\it Run 6} in 2017 April used  for the first time the new iXon-888 CCD
camera,  on an  engineering night  with a  strong wind  and occasional
transparent clouds.

{\it Run 7}  in 2017 May used one allocated  night.  For several hours
the telescope remained closed owing to high humidity.

{\it  Run 8} on  2017 June  6 used  one full  night allocated  for the
multiple-star program. The  sky was mostly clear.  A  record number of
319 targets were observed during this night.

{\it Run 9} in 2017 July was a half-night allocation for the program by
B.~Pantoja. Some observations were made through transparent clouds. 

{\it Run 10}  used the engineering time on  a partially cloudy night,
when other planned tasks could not be accomplished.

{\it  Run   11}  was  almost  a   full  night  of   2017  September  5
(engineering). The regular allocated night of September 11 was lost to
clouds.

{\it Run  12} on  2017 October 4  used only 2  hours of  engineering
time, again observing through the clouds.

{\it Run  13} on  2017 October 28  (half-night) enjoyed a clear  sky and
good seeing. The increased sensitivity of the new camera allowed us to
observe some stars as faint as $I=14$ mag without adaptive correction.
However, SOAR vibrations affected some data of this run. 

\subsection{Data processing}
\label{sec:dat}

Data processing  is described in  TMH10 and subsequent papers  of this
series.  We  recall it here briefly, emphasizing  the caveats. A  series
of  short-exposure images  are recorded  as FITS  cubes,  typically of
200$\times$200$\times$400 pixels size, two cubes per target and per filter. A
larger  image size  of 400$\times$400 pixels  is  used for  pairs wider  than
1\farcs5.  The power spectrum (PS) of each data cube is computed after
subtracting  the bias and applying a threshold  to eliminate  noise  in empty
pixels.   In the  case of  iXon-888 cooled  to $-60^\circ$C,  the dark
current is negligible,  while the bias has a  gradient in the vertical
direction only; the software  was adapted accordingly.  Along with the
PS,  the  program  computes  the  average re-centered  image  and  the
shift-and-add  (SAA)  image  centered  on  the  brightest  pixel.  The
auto-correlation function (ACF) is computed later from the PS, filtered
to remove low spatial frequencies.

Binary companions are detected  in the ACF. Their parameters (position
angle $\theta$, separation $\rho$ and magnitude difference $\Delta m$)
and  their formal  errors are  determined by  fitting a  model  to the
high-frequency part of the PS; the model is a product of the reference
PS and the  PS of two point sources.  In most  cases, the reference is
derived from the azimuthally averaged PS of the object itself (TMH10).
Vibrations,  telescope wind  shake, and  residual aberrations  such as
astigmatism  create two-dimensional patterns  in the  PS that  are not
captured  by  the  model.    Using  another  observed  object  (either
unresolved or with a  substantial magnitude difference) helps here, as
explained  in  \citet{SAM15}.  However,  the  PS structure  constantly
evolves in time, complicating its modeling.

Measurements of  binaries wider than $\sim$0\farcs1  are not sensitive
to the PS model and are very robust. In contrast, for close pairs with
a substantial  magnitude difference ($\Delta  m$), the results  of the
fitting procedure do  depend on the PS structure, the  use (or not) of
the real on-sky reference, and  its conformity to the actual PS.  Some
measures  presented  here  are  affected by  these  poorly  quantified
biases.   Differences between positions  measured in  the $y$  and $I$
filters  are indicative  of such  cases.  Measures  of binaries  at or
below the diffraction limit (27\,mas  in the $y$ filter and 40\,mas in
the  $I$  filter) should  also  be  treated  with caution.    Less
  reliable measurements are marked by colons.

Yet  another  caveat is  related  to  the  differential photometry  of
binaries wider  than $\sim$1\arcsec. The speckle signal  is reduced by
anisoplanatism, biasing  the derived $\Delta m$ to  larger values.  If
the  pair   is  resolved  in  the  centered   images,  an  alternative
photometric procedure  corrects for this bias  (see TMH10).  The
  resolution in  the centered  images, $\beta$, is  also determined  in the
process.  However,  images of wide  pairs can be  partially truncated,
especially  when the wind-induced  telescope shake  causes substantial
image wander or when the seeing is particularly poor.  Such situations
also lead to  an over-estimate of $\Delta m$.  It  is safe to consider
the  published  $\Delta  m$  of  wide  pairs  as  upper  limits.   The
photometry  is  reliable  when  there are  several  mutually  agreeing
measurements of $\Delta m$.

Speckle processing determines the PA of the pair modulo 180\degr. When
the companion is  seen in the SAA images, the  correct quadrant can be
chosen, provided that $\Delta m >  0.3 $ mag; otherwise, the two peaks
in the SAA image are equal and the strongest one cannot be identified.
Quadrants defined in this way are marked by the flag ``q'' in the data
tables.   The  flag ``*''  indicates  binaries    resolved in  the
  centered images, where the quadrants are also known.

\subsection{Calibration of position angle and scale}
\label{sec:cal}

The calibration  of the PA and pixel  scale was done
with  respect  to 64  wide  pairs, as  explained  in  SOAR14.  It  was
revisited and  improved here by including  more calibrators and  the latest
data.   Moreover, the  motion of  some calibrators  is now  modeled by
orbits adjusted to fit the  SOAR data, rather than by linear functions
of time.  A few stars  showing obvious deviations from the models were
removed  from the list  of calibrators.   One such  ex-calibrator that
turned   out  to   be  a   triple   system  is   presented  below   in
Section~\ref{sec:BU83}. A typical rms deviation of the calibrator binaries
from  the models is  from 1  to 3  mas in  both radial  and tangential
directions. Overall,  measures of the calibrators  comprise about 10\%
of all measures presented here.

In run 3, the new SOAR TCS  was used for the first time. We found that
the instrument  PA, nominally  set at 0\degr,  90\degr, or  some other
round  number, was  incorrect  in  a small  number  of cases,  causing
manifestly wrong  angle measurements.   This prompted us  to recompute
all instrument  PAs using the information on  the telescope elevation,
Nasmyth  rotator  angle, and  star  position.   The recomputed  angles
differed from their  nominal round values; they were  used in the data
reduction.  With the recomputed  instrument angles, the rms scatter in
the PA  of the calibration  binaries decreased from 0\fdg7  to 0\fdg4,
indicating  the  appropriateness  of  this correction.   However,  the
remaining scatter  is still substantially  larger than normal.   It is
possible  that the  mechanical  rotation of  the  Nasmyth bearing  had
failures preventing it from reaching the required angle.  This problem
apparently persisted in run 5, but the PAs were not recomputed because
this did not reduce the scatter of the calibrators, 0\fdg4.  The small
engineering run 4 suffered  from the communication problem between the
instrument  software and  the TCS;  some PAs  in these  data  might be
erroneous.  In  contrast, the calibrator  observations in 2017  show a
small PA scatter from 0\fdg1 to 0\fdg2 in all runs.

\begin{figure}
\epsscale{1.1}
\plotone{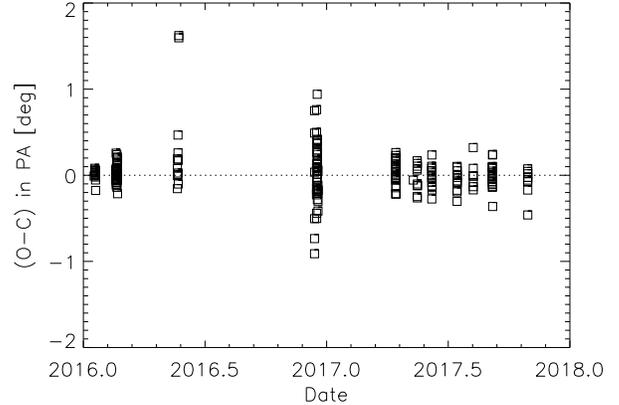}
\caption{Residual in position angle of the calibration stars, plotted vs. time.
\label{fig:calib} }
\end{figure}

Figure~\ref{fig:calib}   plots   the   residual  deviations   of   the
calibrators  in PA  vs.  time.   The  distribution of  points in  time
corresponds to the observing runs in Table~\ref{tab:runs}. Despite the
correction of  the instrument  angles in run  4, there is  one deviant
measure.  The  run~5 also shows a  larger than usual scatter  in the PA
residuals  of  the calibrators.   When  all  231   observations  of  the
calibrators  during two years  are treated  as one  data set,  the rms
scatter in PA is 0\fdg25.  The  global rms scatter of the scale factor
is 0.0041.

Four binaries (WDS J04136+0743, J07277+2127, J09285+0903, and J22409+1433)
have  very  accurate  orbits  based  on  long-baseline  interferometry
\citep{Mut2010b,Mut2010d}. A  total of  10 measures of  those binaries
have  mean  residual   in  PA  of  $-0\fdg15$  with   rms  scatter  of
0\fdg28. The mean  residual in separation is 3.3\,mas  with the rms of
2.0\,mas.  This  comparison  is the  external  check  of  the  data
accuracy. The majority of visual orbits are less accurate than our
measures.  

\subsection{Optical ghosts}
\label{sec:ghosts}

\begin{figure}
\plotone{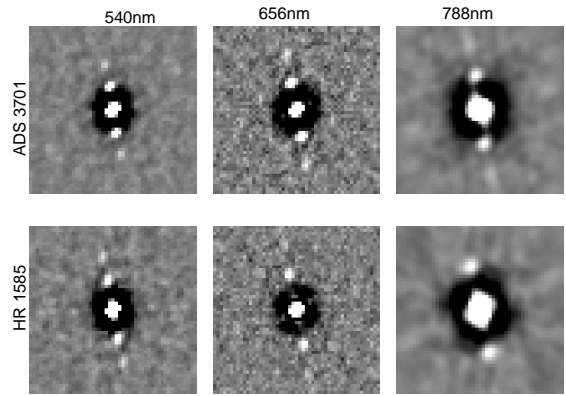} 
\caption{ACFs of ADS 3701 (top row)  and HR 1585 (bottom row) in three
  filters  showing  OGs.  Fragments  of  ACFs  of 51$\times$51  pixels
  (0\farcs78) recorded  on 2016 February  20 are shown.  North  is up,
  East to the right, the intensity scale is arbitrary. 
\label{fig:mosaic} }
\end{figure}

During the first hours of the  2016 February 20 night, a phenomenon of
{\it optical  ghosts} (OG) was observed.  The sky was  clear, the wind
speed was low or zero.  We pointed and unexpectedly resolved ADS~3701,
a  known ``ghost''  binary \citep{Tok2012a}.   However, a  bright star
HR~1585 observed immediately after  also displayed a similar doubling.
Both  bright objects  were  observed with   short 2-ms  exposures,
eliminating     potential    effect     of     telescope    vibration.
Figure~\ref{fig:mosaic}  shows the  speckle 
ACFs  of these  two objects  in three  filters.  Unlike  real double
stars, the separation of the ``companions'' increases in proportion to
the wavelength.  Moreover, we  see the second, fainter companions with
double  separation.   The  phenomenon  is  obviously  associated  with
diffraction  on  a  periodic  structure  where the  first  and  second
diffraction  orders  are  seen.   The separation  of  the  first-order
diffraction  maximum $\rho \sim  0\farcs1$ is  related to  the grating
period $b =  \lambda/\rho$, with $b$ calculated to be  from 0.8 to 1.1
m.

The OGs were observed in three  objects located close to each other on
the sky,  then disappeared  in the following  group of objects  in the
same  sky  area, and  reappeared  again  in  two more  episodes,  each
counting several  successive objects.  The three OG  episodes occurred
during a  time period of about  one hour (from  UT 0:46 to 1:37)  and no
more  OGs were  seen  for the  rest  of the  night.  Data  examination
revealed that  similar OGs  occurred on the  previous night  around UT
0:18 and 9:26. The wind speed on the previous night was also very low.

Table~\ref{tab:OG} provides the  circumstances of some OG observations
that might help in finding their origin.  The first three columns give
the UT  date of  the observation, filter,  and WDS designation  of the
target. The WDS  is used only for convenience, as  OGs are unrelated
to  binary  companions  (in   fact,  04357+0127  is  the  single  star
HIP~21411).  Then  follow the  angle of the  Nasmyth rotator  ROT, the
telescope azimuth  AZ, and the elevation EL.   The difference EL$-$ROT
shows the  instrument angle relative  to the telescope  primary mirror
or, equivalently,  the parallactic angle.   The last two  columns give
the  position angle  $\theta$ and  the  separation $\rho$  of the  OGs
measured as if they were double stars.

\begin{deluxetable*}{l  c  l r r r c   cc }
\tabletypesize{\scriptsize}    
\tablecaption{Examples of optical ghosts
\label{tab:OG} }                    
\tablewidth{0pt}     
\tablehead{ 
\colhead{Date (UT)}  &
\colhead{Filt.}  &
\colhead{WDS}  &
\colhead{ROT}  &
\colhead{AZ}  &
\colhead{EL}  &
\colhead{ROT$-$EL}  &
\colhead{$\theta$}  &
\colhead{$\rho$}  \\
 &  & 
\colhead{ $\alpha$,$\delta$(2000) } & 
\colhead{(\degr)} & 
\colhead{(\degr)} & 
\colhead{(\degr)} & 
\colhead{(\degr)} & 
\colhead{(\degr)} & 
\colhead{(\arcsec)} 
}
\startdata
2016-02-20 00:18   & I  & 04357+0127 & 77.3 & 334.6 & 55.3 & 22.0 & 161.7 & 0.178 \\ 
2016-02-20 09:26   & I  & 14375+0217 & 58.8 & 358.4 & 57.3 & 1.5 & 167.6 & 0.231\\
2016-02-20 09:26   & y  &  14375+0217 & 58.8 & 358.4 & 57.3 & 1.5& 166.2 & 0.153 \\ 
2016-02-21 01:12   & y  & 05348+0929 &  65.9 & 339.6 & 47.9 & 18.1 & 167.5 & 0.155\\
2016-02-21 01:14   & H$\alpha$  & 05348+0929 &  65.9 &  339.6 & 47.9 & 18.1 & 167.0 & 0.184   \\
2016-02-21 01:16   &  y & 05079+0830  &  73.0 & 328.8 & 45.9 & 27.1 & 164.9 & 0.092 \\
2016-02-21 01:30   & y  & 05354$-$0555 & 93.0  & 321.4  & 60.0 & 33.1 & 164.9 & 0.187  
\enddata
\tablecomments{The columns contain: ROT --- Nasmyth rotator angle; AZ
  --- telescope azimuth; EL --- telescope elevation; ROT$-$EL ---
  instrument angle on the sky; $\theta$ and $\rho$ --- PA and
  separation of the OG processed as a binary companion. }
\end{deluxetable*}

The position angles of OGs observed on both nights are confined within
a narrow range, close to  but not exactly aligned with the North-South
direction. The parallactic  angle ranged over 32\degr, so  OGs are not
aligned  in the  vertical direction.  The OGs  were observed  when the
telescope   was  pointing   to  the   North-North-East,   at  moderate
elevation. They  appeared and disappeared  on a time scale  of several
minutes; however, in the pair of data cubes of the same star taken one
after another the OGs are always similar.

The  nature  of OGs  is  mysterious. We  could  reproduce  the OGs  in
simulated speckle images by placing in the beam a fixed periodic phase
screen,   in   addition  to   the   random  atmospheric   perturbation
corresponding to  the 0\farcs7  seeing.  By trial  and error  we found
that  the   clipped  sine  wave  (only   positive  half-periods,  zero
otherwise)  with  a  spatial  period  $b$ of  1\,m  and  a  path-length
amplitude  of 0.2\,$\mu$m  matches the  OGs seen  in  various filters,
creating  two  diffraction orders  of  approximately correct  relative
intensity.

The  origin   of  such   quasi-periodic  phase  disturbances   is  not
known. They are certainly not  related to the instrument or telescope,
as evidenced  by the  position angles. The  fact that OGs  appear only
episodically under zero-wind conditions  suggests that they might be a
transient  atmospheric   phenomenon  like  stratification   or  waves.
Systematic wavefront  distortion under low  wind has been  observed at
the VLT by \citet{SPHERE}.  OGs corresponding to the grating period of
$\sim$2\,m were seen during previous speckle runs at SOAR (see Fig.~11
in  TMH10). However,  OGs with  two  diffraction orders  and a  larger
separation were observed at SOAR for  the first time only now.  The OG
phenomenon can  explain some  false double-star discoveries  made with
speckle interferometry in the past. We mistakenly assumed HIP~75050 to
be a newly  resolved pair before realizing that  its measure refers to
the OG.  


\section{Results}
\label{sec:res}

\subsection{Data tables}

\begin{deluxetable*}{l l l  ccc  rc cc l r r l }                                                                                                                                
\tabletypesize{\tiny}   
\tablecaption{Measurements of double stars at SOAR  (fragment)                                                                                                                              
\label{tab:double} }                                                                                                                                                            
\tablewidth{0pt}                                                                                                                                                                
\tablehead{                                                                                                                                                                     
\colhead{WDS} & \colhead{Discoverer} & \colhead{Other} & \colhead{Epoch} & \colhead{Filt} & \colhead{N} & \colhead{$\theta$} & \colhead{$\rho \sigma_{\theta}$} &               
\colhead{$\rho$} & \colhead{$\sigma \rho$} & \colhead{$\Delta m$} & \colhead{[O$-$C]$_{\theta}$} & \colhead{[O$-$C]$_{\rho}$} & \colhead{Reference} \\         
\colhead{(2000)} & \colhead{Designation} & \colhead{name} & +2000 & & & \colhead{(deg)} & (mas) & ($''$) & (mas) & (mag) & \colhead{(deg)} & \colhead{($''$)}                   
& \colhead{code$^*$}  }   
\startdata     
00024+1047 &	A 1249 AB &	HIP 190 &	17.6802 &	I &	2 &	246.4 &	0.5 &	0.3003 &	0.1 &	0.8 q &	$-$0.1 &	0.041 &	Zir2003  \\ 
00029$-$7436 &	TDS 3 AB &	CD$-$75 1309 &	17.6804 &	I &	2 &	46.8 &	1.4 &	1.5189 &	0.5 &	0.6 *  &  &   &   \\ 
00036$-$3106 &	TOK 686 &	HIP 290 &	16.9487 &	I &	3 &	13.3 &	0.8 &	0.1293 &	12.0 &	3.6    &  &   &   \\ 
            &	       &	           &	17.6801 &	I &	2 &	24.1 &	10.0 &	0.1320 &	1.0 &	4.1    &  &   &   \\ 
00039$-$5750 &	I 700 &	HIP 306 &	16.9596 &	I &	2 &	167.1 &	0.3 &	0.2557 &	0.1 &	0.3    &  &   &   \\ 
00061+0943 &	HDS 7 &	HIP 510 &	17.6802 &	I &	2 &	196.3 &	0.3 &	0.2604 &	0.1 &	0.3   &	0.2 &	0.003 &	FMR2017c  \\ 
00098$-$3347 &	SEE 3 &	HIP 794 &	17.6801 &	I &	2 &	123.6 &	0.1 &	0.9069 &	0.1 &	1.0 * &	1.9 &	0.007 &	Hrt2010a  \\ 
00100+0835 &	A 1801 &	HIP 807 &	17.6802 &	I &	2 &	198.5 &	0.1 &	0.3164 &	0.2 &	0.3    &  &   &   \\ 
00106$-$7313 &	I 43 AB &	HIP 865 &	17.6011 &	I &	3 &	196.0 &	0.7 &	0.4956 &	0.7 &	1.3   &	$-$0.6 &	$-$0.009 &	Cve2010e  \\ 
00121$-$5832 &	RST 4739 &	HIP 975 &	16.9596 &	I &	2 &	282.9 &	0.2 &	0.2853 &	0.1 &	0.2    &  &   &   \\ 
            &	       &	           &	17.6008 &	I &	2 &	279.3 &	1.3 &	0.2776 &	0.5 &	0.0 :  &  &   &   \\ 
00135$-$3650 &	HDS 32 &	HIP 1083 &	16.9487 &	I &	2 &	11.4 &	0.5 &	0.2592 &	0.2 &	0.8 q &	0.0 &	0.000 &	Tok2017b  
 \enddata
\tablenotetext{*}{References to VB6 are provided at  \url{http://ad.usno.navy.mil/wds/orb6/wdsref.txt} } 
\end{deluxetable*}

\begin{deluxetable*}{l l l   c c c  c c c c  }                                                     
\tabletypesize{\scriptsize}               
\tablecaption{Unresolved stars  (fragment)                                                                 
\label{tab:single} }                                                                            
\tablewidth{0pt}                                                                                
\tablehead{                                                                                     
WDS (2000) & \colhead{Discoverer} & \colhead{Hipparcos} & \colhead{Epoch} & \colhead{Filter} &  
\colhead{N} & \colhead{$\rho_{\rm min}$} &    \multicolumn{2}{c}{5$\sigma$ Detection Limit} & $\Delta m$   \\             
$\alpha$, $\delta$ (J2000)  & \colhead{Designation} & \colhead{or other} & \colhead{+2000} & & &                           
& \colhead{$\Delta m (0\farcs15)$}  & \colhead{$\Delta m (1'')$}    & flag    \\             
 & \colhead{or other name}  & \colhead{name}  & & & & (arcsec) & \colhead{(mag)} &  \colhead{(mag)}  &     
}                                                                                               
\startdata     
00219$-$2300 &	RST 5493 A &	HIP 1732 &	17.6801 &	I &	2 &	0.041 &	2.53 &	5.68  \\ 
00291$-$0742 &	MLR 2 &	HIP 2275 &	17.6801 &	I &	2 &	0.041 &	2.66 &	4.41  \\ 
00313$-$1909 &	B 6 &	HD 2797 &	17.6801 &	I &	2 &	0.041 &	3.10 &	4.82  \\ 
00324+0657 &	MCA 1 Aa,Ab &	HIP 2548 &	16.9595 &	I &	2 &	0.040 &	2.93 &	4.61  \\ 
            &	       &	           &	16.9595 &	H$\alpha$ &	2 &	0.041 &	3.32 &	4.24  \\ 
            &	       &	           &	17.6802 &	I &	2 &	0.041 &	2.22 &	4.56  \\ 
            &	       &	           &	17.6802 &	y &	1 &	0.028 &	3.53 &	5.79  \\ 
            &	       &	           &	17.6802 &	H$\alpha$ &	1 &	0.039 &	3.73 &	5.01  \\ 
00366$-$4908 &	HIP 2888 &	HIP 2888 &	16.9596 &	I &	2 &	0.040 &	4.13 &	6.21  \\ 
00374$-$3717 &	I 705 &	HIP 2944 &	16.9596 &	I &	2 &	0.040 &	4.07 &	6.29  \\ 
            &	       &	           &	16.9596 &	y &	2 &	0.028 &	4.39 &	6.05  \\ 
00467$-$0426 &	LSC 10 Aa,Ab &	HIP 3645 &	16.9569 &	I &	2 &	0.040 &	3.14 &	5.10  
\enddata 
\end{deluxetable*}

We do not  report here the results belonging to  other PIs, namely the
Orion and Chamaeleon \citep{Cha}  surveys and the targets observed for
B.~Pantoja.  However, observations of  binaries in  the K2  fields (PI
Ji~Wang) are published here with the PI's permission.

The data tables have the same format as in the previous papers of this
series.   They   are    available   in   full   only   electronically.
Table~\ref{tab:double}  lists 2483  measures of  1570  resolved pairs
 and  subsystems, including 44 newly resolved  pairs. The columns
of Table~\ref{tab:double} contain (1) the WDS \citep{WDS} designation,
(2)  the ``discoverer  designation'' as  adopted  in   WDS, (3)  an
alternative name, mostly from  the {\it Hipparcos} catalog, (4) Julian
year  of observation, (5)  filter, (6)  number of  averaged individual
data  cubes, (7, 8)  position angle  $\theta$ in  degrees  and internal
measurement  error in tangential  direction $\rho  \sigma_{\theta}$ in
mas,  (9,10) separation $\rho$  in arcseconds  and its  internal error
$\sigma_{\rho}$ in mas, and  (11) magnitude difference $\Delta m$.  An
asterisk follows  if $\Delta m$  and the true quadrant  are determined
from the resolved re-centered image; a colon indicates that the data
are  noisy and  $\Delta m$  is  likely over-estimated  (see TMH10  for
details); the  flag ``q'' means  that the quadrant is  determined from
the SAA image (Section~\ref{sec:dat}).  Note that in the cases of multiple
stars,  the positions  and photometry  refer to  the  pairings between
individual stars, not the photo-centers of subsystems.

For  binary stars  with known  orbital elements,  columns  (12--14) of
Table~\ref{tab:double}  list the residuals  to the  ephemeris position
and code  of reference to  the orbit adopted  in the Sixth  Catalog of
Orbits      of     Visual     Binary      Stars     \citep[][hereafter
  VB6]{VB6}.\footnote{See
  \url{http://ad.usno.navy.mil/wds/orb6/wdsref.html}}

Table~\ref{tab:single} contains the data on 426 unresolved stars, some
of  which are  listed as  binaries in  WDS or  resolved here  in other
filters.    Columns   (1)   through   (6)   are   the   same   as   in
Table~\ref{tab:double}, although column  (2) also includes other names
for objects  without discoverer designations.   For stars that  do not
have entries in  WDS, WDS-style codes based on  the J2000 position are
listed in column  (1).  Column (7) is the  estimated resolution limit,
equal to the diffraction  radius $\lambda/D$ for good-quality data and
larger for poor data   (the effective resolution limit is computed
  from the  maximum spatial  frequency where the  signal in  the power
  spectrum stands above the noise, see TMH10).  Columns (8, 9) give the $5
\sigma$  detection limits  $\Delta  m$ at  $0\farcs15$ and  1\arcsec
~separations determined by the procedure described in TMH10.  When two
or  more data cubes  are processed,  the largest  $\Delta m$  value is
listed.   The  last  column   marks  with  colons  noisy  data  mostly
associated  with faint  stars.  In  such cases,  the  quoted detection
limits might be too  large (optimistic); however, the information that
these  stars were  observed  and  no companions  were  found is  still
useful.  In a few  instances, $\Delta m(0\farcs15)=0$ indicates that
the automatic  procedure failed to determine detection  limit at close
separation.

\subsection{Most Tycho binaries are spurious}
\label{sec:Tycho}

The targets in  run 4 featured all WDS binaries  in the {\it Kepler-2}
fields,  including  those  discovered   by  the  {\it  Tycho}  mission
(discoverer codes TDS and TDT).  There are 17 targets with these codes
in our data. One of those, TDT~721, is too wide to be resolved. Of the
remaining  16,  only  one  (TDT~3  at  1\farcs5)  is  confirmed.   The
parameters  of the  {\it Tycho}  pairs make  them easily  accessible to
HRCam.   We  therefore  conclude  that  a large number  of  {\it  Tycho}
binaries are spurious.  Their supposed separations range from 0\farcs4
to 3\arcsec.  Interestingly,  we have resolved two more  pairs in this
group, but  at different  separations: WDS J16086$-$2540  at 1\farcs89
and  $\Delta  I =  6.1$  mag (the  TDS~9771  is  listed with  0\farcs4
separation  and  $\Delta m  =  0.12$  mag)  and WDS  J17022$-$2820  at
0\farcs63 and $\Delta I = 3.8$  mag (TDT~186 is listed at 0\farcs4 and
$\Delta  m  = 0.44$  mag).  These  faint  companions are   random
discoveries unrelated to the previously claimed {\it Tycho} pairs. 

The WDS  contains 14,170 {\it Tycho}   pairs; 330 of those  have a code
'X', i.e. are marked as spurious, while 1201 are confirmed. The
veracity of most {\it Tycho} pairs still waits for confirmation.

\subsection{Newly resolved pairs}
\label{sec:new}

\begin{deluxetable*}{l l l  cc  r r l lr }                                                                                                                                
\tabletypesize{\scriptsize}                                                                                                                                                     
\tablecaption{Newly resolved   pairs
\label{tab:new} }                    
\tablewidth{0pt}  
\tablehead{
\colhead{WDS} & \colhead{Discoverer} & \colhead{Other} & \colhead{Epoch}  &  \colhead{Filt} &\colhead{$\theta$} &                
\colhead{$\rho$}  & \colhead{$\Delta m$} & \colhead{Spectral} &  \colhead{$p$}  \\         
\colhead{(2000)} & \colhead{Designation} & \colhead{name} & +2000 & &
\colhead{(deg)} &  \colhead{($''$)} &  \colhead{(mag)} & \colhead{type} & \colhead{(mas)} }                       
\startdata 
02460$-$0457 &	BU 83 Aa,Ab &	HIP 12912 &	16.9590 &	I &	67.8 &	0.2231 & 4.3    & F3V  & 13.2  \\ 
04274$-$2912 &	TOK 709   &	HIP 20802 &	16.1373 &	I &	68.7 &	0.1278 & 2.1 q  & G6V  & 12.1*   \\ 
06272$-$3706 &	TOK 710   &	HIP 30712 &	16.1330 &	I &	159.4 &	0.1392 & 2.3 q  & ApSrEu & 6.0*  \\ 
07250+0406 &	TOK 711   &	HIP 35986 &	16.1346 &	I &	156.2 &	0.0301 & 0.7    &  G0V   & 6.7*  \\ 
08314$-$6531 &	TOK 712   &	HIP 41800 &	16.1373 &	I &	203.3 &	0.0517 & 1.1 q  &  ApSi   & 4.0  \\ 
09090$-$3802 &	TOK 713   &	HIP 44914 &	16.1375 &	I &	266.5 &	1.3401 & 2.0 *  & K5V    & 14.4*  \\
10476$-$1538 &	TOK 714   Aa,Ab &HIP 52792 &	16.1376 &	I &	72.9 &	0.0294 & 0.1    & F7II/III    & 13.1*   \\ 
11022$-$4230 &	TOK 715   Aa,Ab &HIP 53938 &	16.1376 &	I &	125.0 &	0.1174 & 1.7 q  & ApSrEuCr & 5.2  \\ 
11098+1531 &	TOK 716   &	HIP 54553 &	16.1351 &	I &	248.0 &	2.6180 & 2.0 *  & M0V    & 23.0*  \\ 
11495$-$1636 &	TOK 717   Aa,Ab &HIP 57660 &	16.1351 &	I &	81.3 &	0.0682 & 0.5 q  & K4V    & 19.0*  \\ 
11500$-$5616 &	TOK 718   &	HIP 57698 &	16.1377 &	I &	202.1 &	0.6002 & 3.2    & F3V    & 8.4*   \\ 
11547$-$1401 &	TOK 719   &	HIP 58084 &	16.1351 &	I &	130.6 &	0.0732 & 0.3 :  & K6V    & 13.9*   \\
12479$-$5127 &	TOK 720   &	HIP 62445 &	16.1353 &	I &	176.1 &	0.0485 & 0.7    & G9IVe  & 13.2*  \\ 
12544$-$0932 &	OCC 387 &	HIP 62985 &	16.3890 &	y &	111.7 &	0.2116 & 3.7    & M2III    & 6.0  \\
12547$-$3930 &	TOK 721   &	HIP 63012 &	16.1354 &	I &	320.8 &	0.1647 & 2.9 q  & F7V    & 6.7*  \\ 
12592$-$6256 &	TOK 722   &	HIP 63377 &	16.1351 &	I &	55.1 &	0.4176 & 2.4 q  & G3V    & 11.2  \\ 
14089$-$4328 &	HJ 4653 Aa,Ab &	HIP 69113 &	17.2833 &	I &	95.2 &	0.0555 & 0.1    & B9V     & 6.2   \\ 
14142+1805 &	TOK 723   &	HIP 69549 &	16.1382 &	I &	350.6 &	0.0816 & 2.4    & G0V    & 11.6*  \\ 
14275$-$3527 &	TOK 724   &	HIP 70693 &	16.1354 &	I &	138.5 &	0.0532 & 0.2    & F8V    & 10.8  \\ 
14494$-$5726 &	HDS 2092 BC &	HIP 72492 &	16.1406 &	I &	170.0 &	0.0567 & 0.3    & F5V    & 8.2*  \\ 
15537$-$0429 &	TOK 725   &	HIP 77843 &	16.1410 &	I &	32.8 &	0.1270 & 1.3 q  & F8/G0V    & 12.5  \\ 
16012$-$4632 & SEE 254 Aa,Ab &  HIP 78485 &     17.4324 &       y &     192.7 & 0.0637 & 1.3 :  & F6V    & 9.0 \\  
16086$-$2540 &	TOK 726     & HD 144785   &	16.3876 &	I &	63.7 &	1.8879 & 6.1 *  & G8IV    & \ldots  \\ 
16315$-$3901 &	HDS 2335 Aa,Ab &HIP 80925 &	16.3907 &	I &	68.6 &	0.0724 & 2.0    & K1V    & 44.5*   \\
16385+1240   & TOK 727      &  HIP 81476  &     17.4327 &       I &     176.4 & 0.0876 & 2.1    & G0  & 10.2* \\    
17022$-$2820 &	TOK 728  &	HD 153709 &	16.3876 &	I &	164.4 &	0.6349 &3.8 :  &  A0IV/V    & \ldots  \\ 
17086$-$2650 &	SEE 319 Aa,Ab &	HIP 83878 &	16.3876 &	I &	40.2 &	0.0422 & 0.9    & B9IV  & 3.5   \\ 
17095$-$2612 &	SKF 2521 Aa,Ab & CPD-26 5829 &	16.3876 &	I &	161.0 &	0.1214 & 0.3 :  &  F8   &   \ldots \\ 
17379$-$3752 &	I 247 AC &	HIP 86286 &	17.5345 &	I &	350.5 &	0.1772 & 2.2    &  G8IV   & 21.0  \\ 
18086$-$2752 &	BU 244 Ba,Bb &	HIP 88864 &	16.3877 &	I &	146.1 &	0.1238 & 1.9    & G8III    &  3.5  \\ 
19035$-$2645 &	LDS 5870 Aa,Ab &K214324736 &	16.3882 &	I &	66.8 &	0.5227 &3.9 :  &   G5V?  &   9.4* \\ 
19139$-$2548 &	B 2475 Aa,Ab & HD 179499 &	16.3880 &	I &	118.3 &	0.3894 & 1.7 q  &  F8   & 2.7*   \\ 
19164$-$2521 &	HJ 5101 Aa,Ab &	HD 180132 &	16.3880 &	I &	127.1 &	0.6919 & 4.4    &  B9/A0V   & 2.8*   \\ 
19197$-$2836 &	B 433 AC &	HIP 94985 &	16.3880 &	I &	51.7 &	2.4366 & 3.9    & G1V  & 8.4    \\ 
19231$-$2833 &	RSS 520 Aa,Ab &	HIP 95278 &	16.3880 &	I &	156.6 &	0.3490 & 2.8 :  &  F0   & 1.3  \\
19239$-$2939 &	HJ 5110 Aa,Ab &	CD-29 16082 &	16.3880 &	I &	215.8 &	0.1166 & 0.9 :  &  F4V  & 1.2*  \\ 
19391$-$2811 &	B 444 Aa,Ab & HD 185233 &	16.3880 &	I &	27.3 &	0.1002 & 1.5 :  &   A5III  &   \ldots \\
20100$-$1303 &	TOK 729   &	HIP 99357 &	17.8246 &	I &	277.2 &	1.2429 & 6.1 *  & F5V    & 4.4  \\
21012$-$3511 &  TOK 344 Aa,Ab &  HIP 103735  &  17.6025 &       I &     168.0 & 0.1747 & 1.7    & G3V & 21.5 \\
20212$+$0249 &  TOK 730     &   HIP 103735 &    16.3900 &       I &     143.85 &  0.2269 & 2.3  &  G0   & 2.3   \\ 
21266$-$4604 &  HJ 5267    Aa,Ab &HIP 105879 &  17.6027 &       I &      53.6 & 0.0874 & 1.9    & F7V & 15.8 \\
21278$-$5922 &	TOK 731    &	HIP 105976 &	16.3883 &	I &	27.3 &	0.1002 & 1.5 :  &  F2IV   &  8.0 \\
21357$-$5942 &	TOK 732    &	HIP 106615 &	16.3901 &	I &	185.3 &	0.7033 & 3.3 q  & G0V    & 12.3  \\
23005$-$3345 &	TOK 733    &	HIP 113598 &	17.8248 &	I &	62.9 &	0.0634 & 1.4    & G4V    & 12.3*   
\enddata
\end{deluxetable*}

Table~\ref{tab:new}  lists 44  newly  resolved pairs.   Its format  is
similar to that of  Table~\ref{tab:double}. For some multiple systems,
we  used existing  discoverer  codes and  simply  added new  component
designations.  The last two columns of Table~\ref{tab:new} contain the
spectral type  (as given in  SIMBAD) and the {\it  Hipparcos} parallax
$p$  \citep{HIP2}.   The  {\it  Gaia}  \citep{Gaia}  parallaxes,  when
available, are preferred; they are marked by asterisks.  We comment on
these  objects below.  The  following abbreviations  are used:  PM ---
proper motion, CPM  --- common proper motion, RV  --- radial velocity,
SB1  and  SB2 ---  single-  and  double-lined spectroscopic  binaries.
Orbital periods are estimated from projected separation as
\begin{equation}
 P^* =(\rho /p)^{3/2} M^{-1/2}, 
\end{equation}
where $\rho$ is the angular separation (assumed to equal the semimajor
axis), $p$  is the  parallax, $M$ is  the mass  sum, and $P^*$  is the
period  in years.  Statistical  relation of  these  estimates to  true
periods is  discussed by  \citet{FG67a}.  Data from  the spectroscopic
Geneva-Copenhagen Survey, GCS \citep{N04} are used for some targets.

{\it HIP 20802  (HD 28388)} is a G6V  astrometric binary \citep{MK05},
on  the  California  exoplanet  search  program  \citep{Isaacson2010};
\citet{Nidever2002} found an RV trend. Estimated period: 25 years.

{\it HIP 35986 (HD 58249)} has not been detected as an SB by the GCS,
possibly owing to its fast axial rotation. Its small separation implies
$P^* \sim 10$\,yr. The pair was not measured in run 5 because of the
charge transfer problem, but was resolved again in 2017.3 (run 6) at similar PA and larger
separation. 

{\it HIP 37012 (HD 45698)} is a chemically peculiar Ap star resolved at
0\farcs14,  implying $P^* = 65$ years. 

{\it HIP  41800 (HD 72881)} is  a chemically peculiar  Ap star without
any prior hints on its binarity; $P^* = 30$ yr. The 6 measures show
a large scatter.  However,  without considering  uncertain measures
marked by colons, the position is stable over one year. 

{\it HIP 44914 (CD-37 5499)} is a high-PM K5V dwarf star with only two
references in  SIMBAD. The new  faint component at  1\farcs34 remained
fixed in  one year, hence it is  physical. The period is  long, $P^* =
900$\,yr.

{\it HIP 52792 (HD 93527)} has a CPM companion HIP~52793 at 30\farcs4,
which is itself an SB2 according to the GCS. The new close pair Aa,Ab is
fast ($P^* = 2$ yr), and we indeed see its fast retrograde motion in one year.

{\it HIP 53938 (HD 95699, V360 Vel)} is a chemically peculiar
Ap star with periodic flux variability. The separation has slightly
increased in one year, matching $P^* = 55$ yr. There is a physical
companion at 19\farcs2.  

{\it HIP  54553 (BD+16  2222)} is  a nearby M0V  dwarf GJ~9348  with a
relatively fast PM.  The new companion at 2\farcs6  is confirmed as physical in 2018.

{\it HIP  57600 (HD 102698)} is  another nearby K4V dwarf  known as an
astrometric  (acceleration)  binary  \citep{MK05}. A  fast  retrograde
motion of the new pair is seen, in agreement with $P^* = 5$ yr.  The
wide companion TOK~281 with similar PM listed in  WDS at 407\arcsec
~separation is likely optical. 

{\it  HIP 57698  (HD  102804)}  is on  the  Magellan exoplanet  search
program.   The  newly  found   companion  at  0\farcs60  can  cause  a
measurable RV trend  despite its long $P^* = 350$  yr. The pair
is confirmed as physical by its re-observation in 2017.

{\it HIP 58084 (BD$-$13 3470) } is  a K6V dwarf found here to
be a 0\farcs07  pair with nearly equal components  and fast retrograde
motion, in agreement with the estimated period $P^* = 10$ yr.

{\it HIP  62445 (HD~111170, V940  Cen) } is a  young chromospherically
active G9IV star  in the Sco-Cen association. The  newly resolved pair
has $P^*=  5 $ yr  and shows some  orbital motion in one  year. The
large  discrepancy between  the  {\it Hipparcos}  (7.7\,mas) and  {\it
  Gaia} (13.2\,mas)  parallaxes is presumably caused by  the effect of
orbital motion on the astrometric data reduction.

{\it HIP 62985 ($\psi$~Vir, HR 4902)} is a bright M-type giant with 149
references in SIMBAD. The faint binary companion  detected in 1975 by
lunar occultations (OCC~387 in the WDS) was directly resolved  for
the first time in 2016 at 0\farcs21 and confirmed next year. The
estimated period is $\sim$100 years. 

{\it HIP 63012 (HD 112145) }  is featured in the GCS, but otherwise it
has  attracted  no  interest  so  far. The  new  0\farcs16  pair  with
$P^*=80$\,yr  is also  not interesting  because of  its  slow expected
motion.

{\it HIP 63377 (HD 112636) } has a companion at 0\farcs42. The
resolution is very secure, but no second measure has been taken so
far. Only a slow motion is expected, $P^* = 160$\,yr.

{\it HIP 69113 (HD 123445)}  is a bright B9V member of the Sco-Cen
  association.   It was  pointed instead  of the  fainter pair  DE, at
  5\farcs3 from the component A.  Detection of the new 56\,mas, nearly
  equal  pair Aa,Ab  is secure;  it is  confirmed in  2018,  at closer
  separation. The estimated period is $P^* = 12$ years. 

{\it HIP 69549 (HD 124605) } has a fast PM and an RV of $-$88.9 km~s$^{-1}$;
it is likely metal-poor. Its resolution at 0\farcs08 in 2016.14 is
secure; however, the pair closed down and was unresolved in 2017.37;
$P^* = 13$ yr. 

{\it  HIP 70693 (HD  126620) }  is a  new 0\farcs05  pair with  $P^* =
8$\,yr.  Its  re-observation  shows  decreasing separation  at  nearly
constant PA.

{\it  HIP 72492  (HD  130264) }  is  the {\it  Hipparcos} pair  HDS~2092
revealed here as a new triple  system; we resolve the secondary into a
0\farcs057 pair  BC and observe its  direct motion; $P^*_{\rm BC}  = 20$ yr.
Like  some  other  triple  dwarfs \citep{Tok2018c},  this is  a  ``double  twin'':  the
estimated masses of  B and C are about 0.8  ${\cal M}_\odot$ each, and
their sum  is close to  the mass of  the main component A,  1.6 ${\cal
  M}_\odot$. 

{\it HIP 77843 (HD 142269) } is a new 0\farcs13 pair with a fast motion,
$P^* = 21$\,yr.

{\it HIP  78475 (HD  143235)} is a  0\farcs7 neglected  binary SEE~254
observed as a ``filler'' to the main program. We discovered it to be a
triple system where  the primary component is a  0\farcs064 pair.  The
separation implies $P^*  \sim 12$ yr (we do see  some motion in one
month).  The orbit of  the subsystem  could be  computed now  from the
``wobble''  in  the trajectory  of  the  outer  pair if    accurate
measurements were available. Unfortunately, this is not the case.

{\it J16086$-$2540 (HD 144785)} in the K2 field is resolved at 1\farcs89, $\Delta
I =  6.1$ mag; the 0\farcs4 {\it Tycho} pair TDS~1977 is spurious, 
see  Section~\ref{sec:Tycho}. 

{\it HIP  80925 (HD 148704)},  a nearby K1V  star with fast PM,  is a
spectroscopic binary  with $P =  31.8$ days.  The newly  resolved pair
Aa,Ab has $P^*=1.5$ yr; it  should be detectable by RV variation.  On
the  other  hand,  the  4\arcsec  ~binary  HDS~2335  is  optical;  the
companion B is seen in 2MASS at 9\farcs4, 25\fdg9.

{\it HIP 81476 (HD 150122)} was known to be a binary from its variable
RV and  astrometric acceleration.  It  belongs to the 67-pc  sample of
nearby dwarfs, although  the {\it Gaia} parallax of  10.2\,mas puts it
now  outside 67\,pc.  D.~Latham  (2012, private  communication)
computed  an unpublished  9 yr  spectroscopic orbit.   The  star was
observed at  SOAR and  unresolved in 2014.30,  but now it  is securely
resolved at 0\farcs088  separation. With $\Delta I =  2.1$ mag, double
lines should  be detectable  in the spectrum,  opening the  prospect of
accurate mass measurement.   The star is on the  California program of
planet search \citep{Isaacson2010}.

{\it  J17022$-$2820  (HD  153709)}   is  a  new  pair  with  0\farcs63
separation and  $\Delta I  = 3.8$ mag  (below the  estimated 5$\sigma$
detection  limit of  3.4 mag).   It was  observed as  the  {\it Tycho}
binary   TDT~186,   which   is    revealed   to   be   spurious,   see
Section~\ref{sec:Tycho}.

{\it HIP 83878 (SEE 319)} is a B9IV star with a slow PM. The status of
the known companion at 7\farcs7  is uncertain (it can be optical). The
newly detected subsystem Aa,Ab  at 0\farcs042, confirmed a year later,
has an estimated period of $\sim$20 yr.

{\it CPD$-$26  5829 (SKF 2521)}  is another wide 7\farcs4  binary from
the  K2 program  where  we  found a  subsystem Aa,Ab.  With the  photometric
parallax of 3.2\,mas, the estimated  period of Aa,Ab is over 100
yr; no motion is detected in one year.

{\it HIP 86286} is a known visual binary I~247. Unexpectedly, we found
two companions B  and C of similar brightness  in a nearly equilateral
configuration, both  at 0\farcs2 separation from the  primary. In 1897
the separation was 1\arcsec. It steadily decreased during the 20th century, in
agreement with the long estimated period $P^* = 200$ yr. Presumably,
the secondary  component is a  close binary BC, not  recognized until
now.  This object  resembles the  similar equilateral  triple I~213
discovered at SOAR in 2015 \citep{SAM15}.

{\it HIP 88864 (HD 165732)} is a G8III giant. In addition to the known
2\farcs2 pair BU~244, we measured the subsystem Ba,Bb at
0\farcs12. Re-observation after one year revealed its retrograde
motion, although the parallax of 3.5\,mas implies $P^*_{\rm Ba,Bb} =
120$ yr. 

{\it  EPIC 214324736  (TYC 6881-1560-1)  } is  the 17\farcs4  CPM pair
LDS~5870 with the {\it Gaia}  parallax of 9.4\,mas. The components' PMs
and  photometric distances  match.    WDS lists  a  discordant first
measure in  1960 which,  if true,  would mean that  AB is  optical; we
believe  the first measure   to  be  in error  because  the
configuration  of   AB  remained  fixed  for  15   years  between  its
measurements by  2MASS and {\it Gaia}. The  newly discovered 0\farcs52
subsystem Aa,Ab is also  fixed, proving its physical nature; $P^*_{\rm
  Aa,Ab} = 330$ yr.

{\it  HD  179499}  is  the  known visual pair  B~2475  at  8\farcs2
separation.   The {\it Gaia}  parallax of  the secondary  component is
2.7\,mas, while it gives no parallax for the primary, possibly because
it  was  resolved.  Indeed,  we  discovered Aa,Ab  at  0\farcs39;  its
position did  not change  in a year.  Its estimated orbital  period is
$\sim$1 kyr. However, the F8 spectral type of A and the colors of both
A and B  indicate a photometric parallax of  5\,mas, inconsistent with
the {\it Gaia} parallax. 

{\it HD 180132}  is a 21\farcs4 pair HJ~5101  AB.  Matching {\it Gaia}
parallaxes and PMs of both components leave no doubt that this pair is
physical,  despite  its first  discordant  measure  in  WDS.   The
companion to A  discovered here at 0\farcs69 is  probably too faint to
be detected by {\it Gaia} ($\Delta I = 4.4$ mag).  The period of Aa,Ab
is long, $\sim$2\,kyr; its position is fixed during  one year.
 
{\it  HIP  94985}  is  a  tight  0\farcs2 binary  B~433  to  which  we
discovered a faint ($\Delta I = 3.9$ mag) companion at 2\farcs44. If
the companion were a distant unrelated star with a zero PM, the
separation of AC would have increased by 38\,mas in one year. Instead,
we found it to be constant to within 3\,mas. The 0\fdg4 change in the
PA during this period is likely related to the angle calibration
problems in run 3. 

{\it HIP  95278} is  a 9\farcs7 pair  RSS~520 located in  a relatively
crowded region  of the  sky; its faint  secondary component  is likely
optical.   The small  PM  and parallax  do  not help  in defining  its
status. The newly discovered subsystem Aa,Ab at 0\farcs39 has a better
chance to be  physical, but its long period of  $\sim$2.5 kyr does not
inspire any interest in this discovery.

{\it CD$-$29 16082 (HJ~5110)} is a 5\farcs9 near-equal pair where we
found a 0\farcs12 subsystem Aa,Ab. This triple system could be
physical. However, the {\it Gaia} parallax of 1.2\,mas implies very
long periods;  if the parallax is correct, both components A and B are located above the main sequence.

{\it  HD 185233}  is a  1\arcsec ~binary  B~444  without trigonometric
parallax or  orbit. We discovered the 0\farcs10  subsystem Aa,Ab. This
resolution,  although not  checked in  2017, is  secure.  The expected
period  of Aa,Ab  is  a  couple of  centuries,  given the  photometric
parallax of 1.9\,mas.

{\it HIP 99357 (HD 191365)  } is an F5V spectroscopic binary according
to the GCS,  based on two mutually discordant measures  of the RV. The new
faint companion at 1\farcs34 with $\Delta I = 6.1$ mag may be optical.

{\it HIP  100355} was  pointed in 2015  by mistake instead  of HLD~158
(WDS  J20213+0250) and  resolved at  0\farcs2, in  marked disagreement
with  the  orbit  of  HLD~158  \citep{SAM15}.  Now  the  confusion  is
clarified, as both HIP~100355 and HLD~158 were measured. Our published
2015 measure should be attributed to HIP~100355.

{\it  HIP 103735  (HD  199918) }  is a  nearby  G3V star  with a  wide
(186\arcsec) CPM companion, which possibly is a white dwarf. The main star is a
spectroscopic   and  astrometric  binary,   first  resolved   here  at
0\farcs17. The separation implies an orbital period of the order of 20
yr  that might  explain  the non-resolution of  this  binary at  Gemini
in 2011 \citep{Tok2012b}.

{\it HIP 105879 (HD 203934)} has astrometric acceleration and variable
RV  (CGS). Its  spectrum has  double  lines, according  to the  ongoing
monitoring at the CTIO 1.5 m telescope.  The first attempts to resolve
the  pair  at Gemini  \citep{Tok2013}  and  at  SOAR in  2015.74  were
unsuccessful,  but in 2017.6  it was  securely resolved  at 0\farcs087.
The  separation   and  magnitude  difference   match  the  preliminary
spectroscopic  orbit with  $P= 5$  yr and  the mass ratio  $q  = 0.78$.
According  to this orbit,  the separation  in 2015.74  was 0\farcs025.
Observations for several more years will lead to accurate measurements
of  the  masses  of components  Aa  and  Ab;  the next  periastron  is
predicted  in 2020.6.   The CPM  companion D  at 44\arcsec  ~makes this
system  triple. The  star D  has a  constant RV  and it  has  not been
resolved at SOAR in 2014.

{\it HIP 105976  (HD 203970) } is resolved at  0\farcs09 in 2016.4; it
moved by 8\degr ~in one year,  matching the estimated period $P^* = 30$
yr. 

{\it HIP 106615 (HD 205158) } has a new faint companion at 0\farcs70  with
$P^*= 300$ yr. The star is on the Magellan planet search program.

{\it  HIP 113598  (HD  217344) }  is  a 3\farcs9  physical pair  B~582
containing  a  1.6  day   SB1  TZ~PsA.   We  tentatively  resolved  an
intermediate subsystem Aa,Ab at 0\farcs063. However, this result needs
confirmation because telescope vibrations could mimic the resolution.
If this pair is real, its period is short, $P^* = 8$ yr.

\subsection{The triple system BU~83}
\label{sec:BU83}

\begin{figure}
\epsscale{1.0}
\plotone{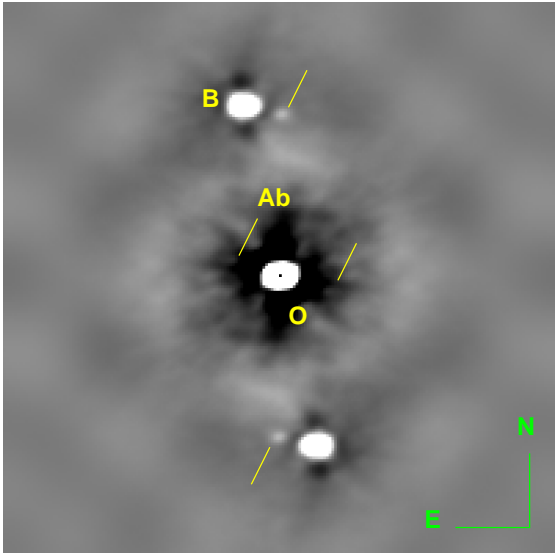}
\caption{ACF  of BU~83  recorded  on  2017.83 in  the  $I$ band.   The
  coordinate  center is  at  O,  the letters  indicate  the two  peaks
  corresponding to B and Ab, while lines mark all four peaks caused by
  the  faint component  Ab (the two  central peaks  are lost  in the
  noise).
\label{fig:BU83} }
\end{figure}

\begin{figure}
\epsscale{1.1}
\plotone{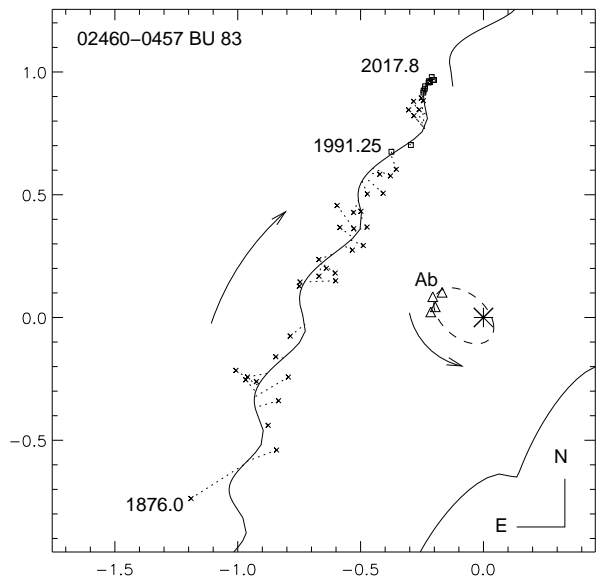}
\caption{Trajectory of the outer subsystem in BU~83 accounting for the
  wobble  caused  by  the  inner  subsystem.   Crosses  ---  micrometer
  measures and  speckle interferometry at small  apertures, squares ---
  accurate  recent  measures  at  SOAR  and  {\it Hipparcos},  line  ---  the
  orbit.  The  dashed  ellipse  and  triangles denote  the  orbit  and
  resolved  measures of  the subsystem  Aa,Ab.  The axis  scale is  in
  arcseconds.
\label{fig:BU83orbit} }
\end{figure}

One of  the calibrator  binaries, WDS J02460$-$0457  (BU~83, HIP~12912,
ADS 2111, F3V),  is found to be triple.  Looking  back at the archival
SOAR data,  the tertiary  component can be  noted in  several $I$-band
ACFs.  The  clearest previous detection was in  2012.92, overlooked at
the  time.  The  inner  pair  Aa,Ab was  then  at 58\fdg7,  0\farcs20,
$\Delta  I =  4.5$  mag; it  turned by  23\degr  ~in 5  years.  A  very
tentative  measure of  Aa,Ab in  2009.671 is  58\fdg9,  0\farcs19.  In
2015,  the  companion's detection  was  marginal,  but  it was resolved
securely in  2016 and 2017 (Figure~\ref{fig:BU83}).   The companion is
not seen in the $y$ band.   Owing to the faintness of the companion Ab,
the fitted triple-star  models do not converge well,  resulting in the
low accuracy of the measures of Aa,Ab.

Interestingly,  the  existence  of  the  subsystem  was  suspected  by
\citet{Dommanget1972}  from the  run  of the  residuals  of the  outer
pair.  He proposed  a 36 yr  astrometric  orbit with  an amplitude  of
0\farcs08. 

Figure~\ref{fig:BU83orbit}  shows  that  the  accurate  SOAR  measures
indeed  deviate from  the long-period  orbit (which  itself  is poorly
defined  by  the short  observed  arc).   Using  the {\tt  orbit3.pro}
code\footnote{           \url{http://dx.doi.org/10.5281/zenodo.321854}}
\citep{orbit3}, we  can fit  the wobble and  the resolved  measures of
Aa,Ab by two  sets of orbital elements.  However,  the low accuracy of
historic measures and the partial coverage of the inner resolved orbit
make such fits  rather uncertain.  The plots in  the Figure correspond
to the  inner period of  38 yr and  the outer period of  915 yr.
The  inner subsystem  has  direct  motion, while  the  outer orbit  is
retrograde, so the angle between  the inner and outer angular momentum
vectors is large.  The sign  of the wobble confirms that the subsystem
belongs to the primary component.  Patient accumulation of measures is
needed  before attempting  a meaningful  analysis of  this interesting
triple  system.   The intuition  of  J.~Dommanget  who discovered  the
wobble in the noisy measures available prior to 1972 is truly amazing,
but the  astrometric orbit derived from  the old data  alone cannot be
trusted.





\acknowledgments 

We thank the  operators of SOAR \fbox{D.~Maturana} (he  passed away in
2017), P.~Ugarte, S.~Pizarro,  J.~Espinoza, C.~Corco, R.~Hernandez for
efficient support of our program, R.~Cantarutti for adapting the HRCam
software to the new detector,  and N.~Law for offering us his iXon-888
camera.  Some observations were made using the Luca-R cameras borrowed
from G.~Cecil and from the STELES  team, and we are thankful for this.
 Detailed  comments by  the Referee helped us  to improve
  the presentation and the tables.

R.A.M. acknowledges  support from the Chilean Centro  de Excelencia en
Astrof\'{i}sica y Tecnolog\'{i}as Afines  (CATA) BASAL PFB/06, the Project
IC120009 Millennium Institute of  Astrophysics (MAS) of the Iniciativa
Cient\'{i}fica Milenio  del Ministerio de  Econom\'{i}a, Fomento y  Turismo de
Chile, and CONICYT/FONDECYT Grant Nr. 117 0854. 

This work  used the  SIMBAD service operated  by Centre  des Donn\'ees
Stellaires  (Strasbourg, France),  bibliographic  references from  the
Astrophysics Data  System maintained  by SAO/NASA, and  the Washington
Double  Star  Catalog  maintained  at  USNO.

{\it Facilities:}  \facility{SOAR}.




\begin{thebibliography}{99}

\bibitem[Brice\~no \& Tokovinin(2017)]{Cha}
Brice\~no, C. \& Tokovinin, A. 2017, AJ, 154, 195


\bibitem[Dommanget(1972)]{Dommanget1972}
Dommanget, J. 1972,  Bull. RAS  Belg., 8,  63



\bibitem [Gaia collaboration(2016)]{Gaia}
Gaia Collaboration, Brown, A. G. A., Vallenari, A., Prusti, T. et
al. 2016, A\&A, 595, 2


\bibitem[Gomez et al.(2016)]{Gomez2016}
Gomez, J., Docobo, J. A., Campo, P. P., \& Mendez, R.  2016, AJ, 152, 216


\bibitem[Hartkopf, Mason \& Worley (2001)]{VB6} 
Hartkopf, W. I., Mason, B. D. \& Worley, C. E. 2001, AJ, 122, 3472 



\bibitem[Hartkopf et al.(2012)]{Hrt2012a} 
Hartkopf, W. I., Tokovinin, A.  \& Mason, B. D.  2012, AJ, 143, 42




\bibitem[Horch et al.(2015)]{Horch2015}
Horch, E. P., van Belle, G. T., Davidson, J. W., Jr. et al. 2015, AJ, 150, 151

\bibitem[Horch et al.(2017)]{Horch2017}
Horch, E. P., Casetti-Dinescu, D. I., Camarata, M. A. et al. 2017, AJ, 153, 212

\bibitem[Isaacson \& Fischer(2010)]{Isaacson2010}
Isaacson, H. \& Fischer, D. 2010, ApJ, 725, 875




\bibitem[Makarov \& Kaplan(2005)]{MK05}
Makarov, V. V. \& Kaplan, G. H.,  2005, AJ, 129, 2420 


\bibitem[Mason et al.(2001)]{WDS}
Mason, B. D., Wycoff, G. L., Hartkopf, W. I. et al.  2001, AJ, 122, 3466 (WDS)

\bibitem[Mason et al.(2018)]{Mason2018}
Mason, B. D., Hartkopf, W. I., Miles, K. N., Subasavage, J. P., Raghavan, D. \& Henry, T. J. 2018, AJ (in press)


\bibitem[Mendez et al.(2017)]{Mdz2017} 
Mendez, R.~A., Claveria, R.~M., Orchard, M.~E., \& Silva, J.~F.\ 2017, AJ, 154, 187 


\bibitem[Muterspaugh et al.(2010a)]{Mut2010b}
 Muterspaugh, M. W., Hartkopf, W. I.,  Lane. B. F. et al. 2010a, AJ, 140, 1623

\bibitem[Muterspaugh et al.(2010b)]{Mut2010d}
Muterspaugh, M. W., Fekel, F. C., Lane, B. F. et al. 2010b, AJ, 140, 1646

\bibitem[Nidever et al.(2002)]{Nidever2002}
 Nidever, D.L., Marcy, G.W., Butler, R.P. et al. 2002 ApJS, 141, 503 

\bibitem[Nordstr\"om et al.(2004)]{N04}
Nordstr\"om,  B., Mayor,  M.,  Andersen, J.  et al.  2004, A\&A, 418, 989  (GCS)
%


\bibitem[Sauvage et al.(2015)]{SPHERE}
Sauvage, J.-F., Fusco, T., Guesalaga, A. et al. 2015, in Proceedings of the Conference 
``Adaptive Optics for Extremely Large Telescopes 4.'' \url{http://www.escholarship.org/uc/item/910646qf}


\bibitem[Tokovinin \& Cantarutti(2008)]{TC08}
Tokovinin, A. \& Cantarutti, R. 2008, PASP, 120, 170 


\bibitem[Tokovinin, Mason, \& Hartkopf(2010a)]{TMH10}
Tokovinin, A., Mason, B. D., \& Hartkopf, W. I. 2010a,  AJ, 139, 743 (TMH10)

\bibitem[Tokovinin et al.(2010b)]{SAM09}
Tokovinin, A. Cantarutti, R., Tighe, R. et al. 2010b, PASP, 122,  1483 

\bibitem[Tokovinin(2012)]{Tok2012a} 
Tokovinin, A.  2012, AJ, 144, 56

\bibitem[Tokovinin et al.(2012)]{Tok2012b} 
Tokovinin, A. Hartung, M., Hayward, Th. L., \& Makarov, V. V.  2012, AJ, 144, 7  

\bibitem[Tokovinin et al.(2013)]{Tok2013}
Tokovinin, A. Hartung, M., Hayward, Th. L. 2013, AJ, 146, 8

\bibitem[Tokovinin(2014)]{FG67a}
Tokovinin, A. 2014, AJ, 147, 86


\bibitem[Tokovinin et al.(2014)]{TMH14}
Tokovinin, A., Mason, B. D., \& Hartkopf, W. I.  2014, AJ, 147, 123 


\bibitem[Tokovinin et al.(2015)]{TMH15}
Tokovinin, A., Mason, B. D.,  Hartkopf, W. I. et al. 2015, AJ, 150, 50  
(SOAR14)

\bibitem[Tokovinin et al.(2016a)]{SAM15}
Tokovinin, A., Mason, B. D.,  Hartkopf, W. I. et al. 2016a, AJ, 152,
116 

\bibitem[Tokovinin et al.(2016a)]{SAM}
Tokovinin A., Cantarutti, R., Tighe R.,  et al. 2016b, PASP, 128, 125003

\bibitem[Tokovinin(2016)]{Tok2016c}
Tokovinin, A.  2016, AJ, 152, 138

\bibitem[Tokovinin \& Latham(2017)]{orbit3}
Tokovinin, A. \& Latham, D. W.  2017, ApJ, 838, 54

\bibitem[Tokovinin(2017)]{Tok2017c}
Tokovinin, A. 2017, AJ, 154, 110

\bibitem[Tokovinin(2018a)]{Tok2018a}
Tokovinin, A. 2018, PASP, 130, 5002

\bibitem[Tokovinin(2018b)]{Tok2018b}
Tokovinin, A. 2018a, Inf. Circ. 194, 1 

\bibitem[Tokovinin(2018c)]{Tok2018c}
Tokovinin, A. 2018c, AJ, in press (DOI https://doi.org/10.3847/1538-3881/aab102)


\bibitem[van Leeuwen(2007)]{HIP2} 
van Leeuwen, F. 2007, A\&A, 474, 653 


\bibitem[ ()]{}




\end{thebibliography}
\end{document}